\documentstyle[12pt,a4,epsf,moriond]{article}

\newcommand{\HI}{H\thinspace\protect\footnotesize I\protect\normalsize}

\newcommand{\ij}{\mbox{$I-J$}}
\newcommand{\jk}{\mbox{$J-K$}}
\newcommand{\B}{{$B$}}

\newcommand{\II}{{$I$}}
\newcommand{\J}{{$J$}}
\newcommand{\K}{{$K_s$}}
\newcommand{\tfr}{Tully\,--\,Fisher relation}
\newcommand{\ccd}{colour\,--\,colour diagram}
\newcommand{\kms}{\,km\,s$^{-1}$}
\newcommand{\etal}{{\it et~al.}}
\newcommand{\cf}{{\it cf.\,}}
\newcommand{\eg}{{\it e.g.},\ }         
\newcommand{\ie}{{\it i.e.},\ }         

\def\la{\mathrel{\hbox{\rlap{\hbox{\lower4pt\hbox{$\sim$}}}\hbox{$<$}}}}
\def\ga{\mathrel{\hbox{\rlap{\hbox{\lower4pt\hbox{$\sim$}}}\hbox{$>$}}}}

\def\deg{{^\circ}}
\def\arcmin{\hbox{$^\prime$}}

\def\fm{\hbox{$.\!\!^{\rm m}$}}

\def\farcm{\hbox{$.\mkern-4mu^\prime$}}

\begin{document}

\heading{DENIS Galaxies in the Zone of Avoidance}

\author{A. Schr\"oder $^{1}$, R.C. Kraan-Korteweg $^{2}$, G.A. Mamon $^{3,2}$,
S. Ruphy $^{4}$} 
{$^{1}$ Institute of Astronomy, NCU, Chung-Li, Taiwan}
{$^{2}$ DAEC, Observatoire de Paris, Meudon, France}
{$^{3}$ IAP, Paris, France}
{$^{4}$ DESPA, Observatoire de Paris, Meudon, France}

\begin{moriondabstract}
We investigated the potential of using DENIS
for studies of
galaxies behind the obscuration layer of our Milky Way, and mapping the
Galactic extinction.
%
As a pilot study, we examined DENIS \II -, \J -, and \K -band images of heavily
obscured galaxies from a deep optical (\B -band) galaxy survey in the Zone of
Avoidance. We tried to uncover additional galaxies at latitudes where the Milky
Way remains fully opaque, \ie we conducted a `blind' search at $|b| < 5\deg$
and $A_B \ga 
4-5^{\rm m}$. Furthermore, we determined the \II , \J\ and \K\ magnitudes of
galaxies in the low-latitude, nearby, rich cluster Abell 3627 and compared the
resulting \ccd\ with that of an unobscured cluster.

\end{moriondabstract}

\section{Introduction }
About 25\% of the optically visible extragalactic sky is obscured by the dust
and stars of our Milky Way. Dynamically important structures -- individual
near-by galaxies as well as large clusters and superclusters -- might still lie
hidden in this zone (\cf\ \cite{Dw1}, \cite{A3627}). Complete whole-sky mapping
of the galaxy and mass distribution is essential in explaining the origin of
the 
peculiar velocity of the Local Group, the dipole in the Cosmic Microwave
Background (CMB), and other large-scale streaming motions.

Various approaches are presently being employed to uncover the galaxy
distribution in the Zone of Avoidance (ZOA): deep optical searches,
far-infrared 
(FIR) surveys (\eg IRAS), and blind \HI\ searches. All methods produce new
results, 
but all suffer from (different) limitations and selection effects. Here, the
near infrared (NIR) surveys such as DENIS \cite{den,denmes} and 2MASS
\cite{2m} 
could provide important 
complementary data. NIR surveys will:

(i) be sensitive to early-type galaxies -- tracers of massive groups and
clusters -- which are missed in IRAS and \HI\ surveys,

(ii) have less confusion with galactic objects compared to FIR surveys, and

(iii) be less affected by absorption than optical surveys.

\noindent But can we detect galaxies and obtain accurate magnitudes in crowded
regions and at high foreground extinction using the DENIS survey?

To assess the performance of the DENIS survey at low galactic latitudes we
compared DENIS data with results from a deep optical survey in the southern ZOA
(\cite{KKW,KKFB,A3627} and references therein). We investigated two regions
in the Great Attractor (GA) area where the galaxy density is high and the
galactic extinction well determined 
\cite{Sey}.  We addressed the following 3 questions:

$\bullet$ How many galaxies visible in the $B_J$-band ($B_{lim} \approx 19.0$)
  can we recover in \II ($0.8\mu m$), \J ($1.25\mu m$) and \K
  ($2.15\mu m$)? Although less affected by extinction (45\%, 21\% and 9\% as
  compared to $B_J$), their respective completeness limits are lower ($17\fm0,
  14\fm5$, and $12\fm2$ \cite{gam1}).

$\bullet$ Can we identify galaxies at high extinction ($A_B > 4-5^{\rm
  m}$) where optical surveys fail and FIR surveys are plagued by confusion?

$\bullet$ Can we map the galactic extinction from NIR colours of galaxies
  behind the Milky Way?

\section {First Results}

\begin{figure} [ht]
\vspace{-.3cm}
\hfil \epsfxsize 14cm
\epsfbox{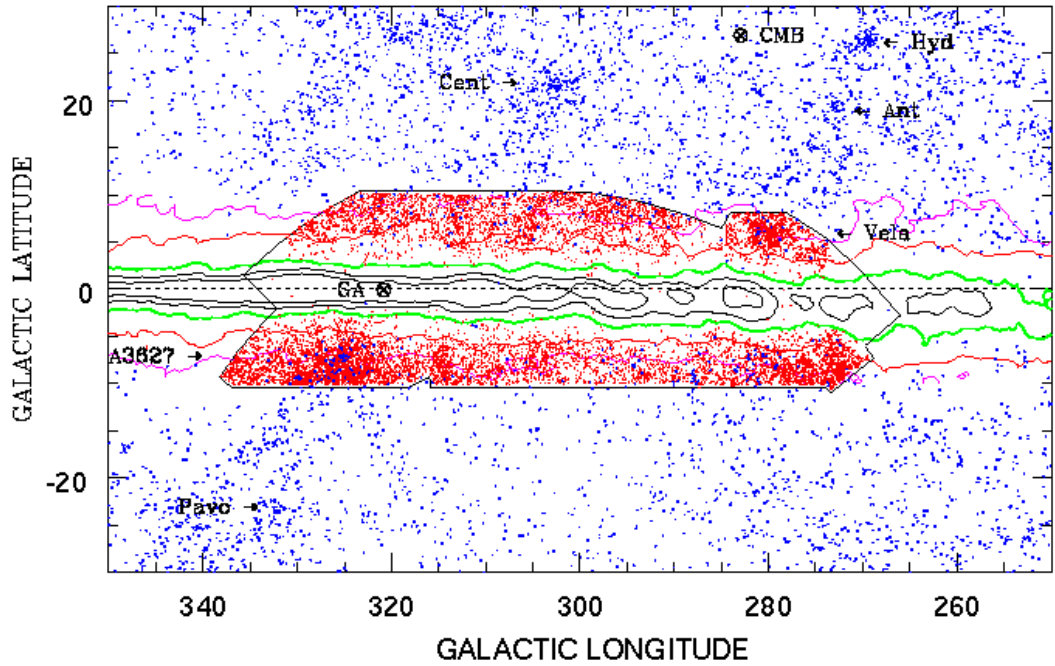} \hfil
\caption{Distribution of Lauberts galaxies (larger dots) and galaxies found by
  Kraan-Korteweg and Woudt in the outlined area (smaller dots). Assuming
  constant gas/dust ratio, the superimposed contours represent absorption
  levels 
  of $A_B=1.5$, 2.5, 5.0 (thick line), 7.5 and 10.0.}
\label{kkwplot}
\end{figure}

Kraan-Korteweg and collaborators have started a deep search for galaxies in the
southern Milky Way using the IIIaJ film copies of the ESO/SRC survey. All
galaxies above the diameter limit of $D=0\farcm2$ and --- depending on surface
brightness --- $B \la 19.0-19.5$ are being catalogued.  Many of the
faint low-latitudes galaxies will be intrinsically bright galaxies. 
In the surveyed area ($265\deg \la \ell \la 340\deg,\|b| \la 10\deg$)
over 11\,000 unknown galaxies were identified.  Their
distribution is shown in Fig.~\ref{kkwplot} together with the larger Lauberts
galaxies ($D \ge 1\farcm0$) \cite{LAU}. The most important
extragalactic structures, the apex of the CMB dipole
and the predicted center of the GA are labelled. Superimposed are
contours of foreground
extinction as determined from \HI -column densities \cite{BH,KKW,A3627}. The
deep 
optical galaxy search clearly reduces the width of the ZOA and reveals many
previously unrecognised extragalactic filaments, clusters and overdensities
({\it e.g.\/,} 
the rich cluster A3627 \cite{A3627}).
The innermost part of the Milky Way, however, remains opaque.

\subsection {Expectation from DENIS}

\begin{figure} [ht]
\centerline {\epsfxsize=10cm \epsfbox[20 161 564 532]{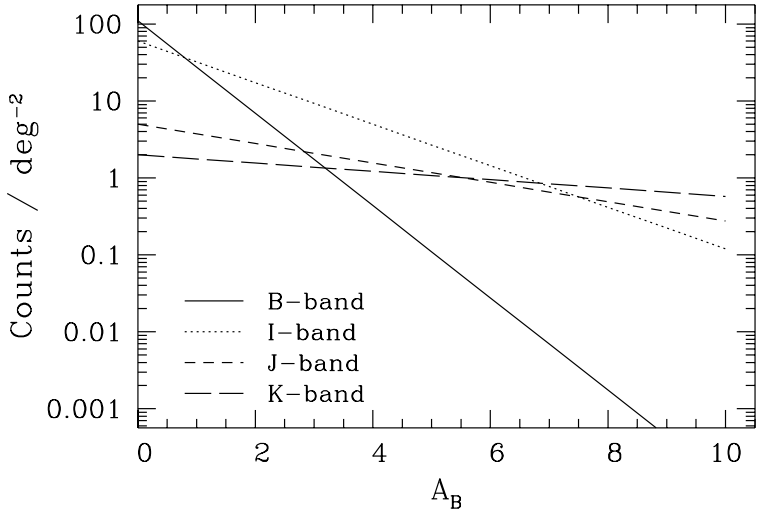}} 
\caption{Galaxy counts in \B , \II , \J\ and \K\ as a function of absorption
in \B . } 
\label{galctsplot}
\end{figure}

Can DENIS do better? Figure~\ref{galctsplot} shows the average number of
galaxies per square degree expected to be found in the respective passbands
as a 
function of galactic foreground extinction.  In unobscured regions, the number
density of galaxies per square degree is 110 in the blue for $B_J\le19.0$
\cite{gar}, and 60, 5, and 2 in the \II , \J\ and \K\ passbands for their
respective completeness limits of $I_{lim}$=17.0, $J_{lim}$=14.5, and
$K'_{lim}$=12.2 \cite{gam1}.  The number counts in the blue decrease with
increasing obscuration as $N(<B) \simeq 110 \cdot
10^{(0.6[B-19])}\,$deg$^{-2}$.  
According to Cardelli \etal\ \cite{Car} the extinction in the NIR passbands are
$A_I$=$0\fm45$, $A_J$=$0\fm21$, and $A_K$=$0\fm09$ for $A_B$=$1\fm0$, hence the
decrease in number counts as a function of extinction is considerably slower.

The NIR becomes notably more efficient at $A_B\simeq 3^{\rm m}$ while the
Milky Way becomes opaque at $A_B \ge 4^{\rm m}$. At an extinction
of $A_B \simeq 6-7^{\rm m}$ we expect to find 1 gal/deg$^2$ in all three NIR
passbands and \K\ becomes superior only at even higher extinction. In April
1997, 
a new cooling system for the focal instrument will be mounted. This
should result in an 
increase of the limiting magnitude in \K\ of 1\,mag, lifting the \K -curve by
0.6 
and therewith the expectation of discovering new galaxies deep in the
obscuration layer of the Milky Way.

These are very rough predictions and do not take into account any dependence on
morphological type, surface brightness, orientation and crowding, which will
surely lower the counts of actually detectable galaxies counts \cite{gam}.

\subsection{Recovery of galaxies found in the \B -band }

To test the probability of re-detecting galaxies found in the optical survey on
the DENIS \II -, \J -, and \K -images, we have selected the dense area of the
Norma cluster (A3627 at $\ell=325\deg$, \mbox{$b=-7\deg$}, \cf\
Fig.~\ref{kkwplot}). 
Three high-quality DENIS strips cross the cluster practically through its
center. On these three strips we inspected 66 images which cover about
one-eighth of the cluster area within its Abell-radius (each image is
$12\arcmin$x$12\arcmin$, offset by $10\arcmin$ in declination). The extinction
over the regarded cluster area varies as $1\fm2 \le$ A$_B \le 2\fm0$.

On the 66 images, 156 galaxies had been identified in the optical survey. We
have 
recovered 125 galaxies in the \II -band, 102 in the \J -band, and 75 in the \K
-band. As suggested in Fig.~\ref{galctsplot}, the \K -band indeed is not
optimal 
for identifying obscured galaxies due to its low magnitude limit.  Most of the
galaxies not re-discovered in \K\ are spiral galaxies, probably because of
their 
lower surface brightness. Surprisingly enough, the \J -band images are found to
be optimal at these latitudes and not the \II -band images. In the latter the
severe star crowding makes identification of faint galaxies very difficult. At
these extinction levels, the optical survey does remain the most efficient in
{\it identifying} obscured galaxies.

\begin{figure} [ht]
\centerline {\epsfxsize=9.cm \epsfbox{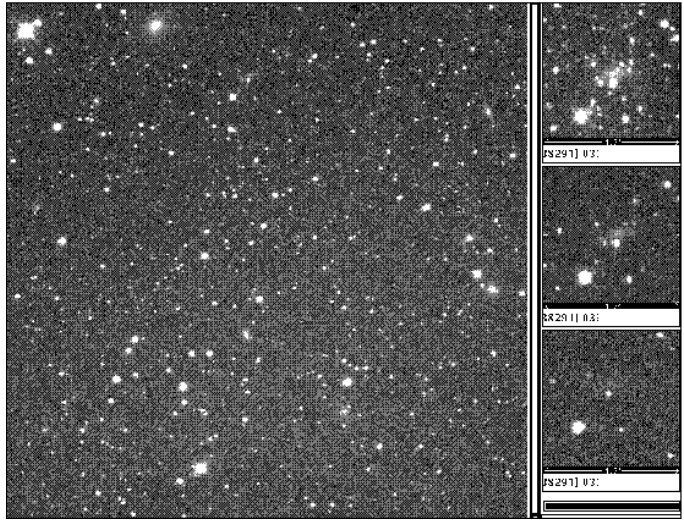}} 
\caption{A DENIS \J -band image of the center of the Norma cluster. The
right-hand shows the \II -, \J - and \K -band image of a zoomed-in spiral
galaxy.} 
\label{norplot}
\end{figure}
Figure~\ref{norplot} shows a DENIS \J -band image of the center of the A3627
cluster, with the cD galaxy at the top. The right-hand side shows the \II ,
\J , 
and \K\ (top to bottom) of a zoomed-in spiral galaxy (top center on large
image).

\subsection{`Blind' search for galaxies}

\begin{figure} [ht]
\hfil{\epsfxsize=8.5cm \epsfbox{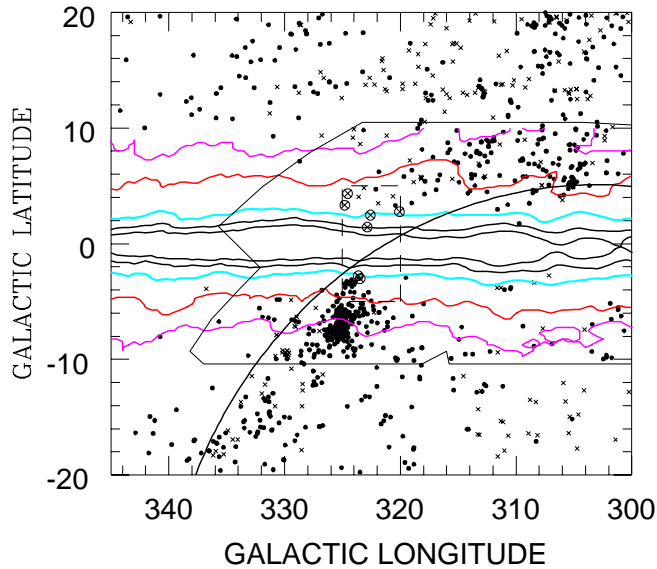}}\hfil
\caption{Distribution of galaxies in the GA region with v$<$7000\kms
(crosses: 0-3000\kms , filled dots: 3000-7000\kms ). The DENIS search area
is marked. Newly found DENIS galaxies are shown as large encircled crosses.}
\label{bsrchplot}
\end{figure}

For the `blind' search we have chosen an area where a nearby filament of
galaxies is suspected to cross the galactic plane from the Norma cluster in the
south to the Centaurus cluster in the north, see Fig.~\ref{bsrchplot}, 
where all galaxies with $v< 7000$\kms\ are shown.  We therefore have a
relatively high probability of finding obscured galaxies. The
search area was defined as
$320\deg \leq \ell \leq 325^\deg$ and $|b|\leq 5\deg$.

Of the 1800 images in this area we have inspected 237 by eye. From the optical
survey 16 galaxies at the high latitude border of the `blind' search area were
known; 13 of these could be re-identified in the \II - as well as in the \J
-band, and 11 in the \K -band. In addition, we found 7 new galaxies in \II\ and
\J , 5 of which also appear in the \K -band. They are plotted as encircled
crosses in Fig.~\ref{bsrchplot}. 
The ratios of galaxies found in \B\ compared to \II , and of \II\ compared to
\K\ are
larger than for the Norma cluster. This obviously is due to the higher
foreground extinction (starting with A$_B \ge 2\fm3 -3\fm1$ at the border
of the search area $|b|=\pm 5\deg$, \cf\ contours in Fig.~\ref{bsrchplot}).

On average, we have found about two galaxies per square degree in the 
\II-band, 
which agrees roughly with the predictions above. The numbers of inspected
images and 
detected galaxies are, of course, too low to allow a statistical conclusion and
compare it to the predictions in Fig.~\ref{galctsplot}. We did look on the one
hand in an overdense region where we expect {\it a priori} more galaxies, on
the other 
hand, at this longitude range we do not expect to find galaxies at latitudes
below $b \le 1\deg-2\deg$ \cite{gam}. The visual impression of the low-latitude
images substantiate this -- the images are nearly fully covered with stars.

The most important result from the test search: highly obscured, optically
invisible galaxies can indeed be unveiled in the NIR.  The lowest galactic
latitude at which we have found a galaxy is $b \simeq 1.5\deg$ with $A_B
\simeq 7\fm5$ 
as estimated from \HI -column densities. 

\subsection{Photometry of galaxies in the Norma cluster }

We have used a preliminary galaxy pipeline (Mamon, in these proceedings), 
on the
Norma cluster identified in section 2.2 to obtain preliminary \II , \J\ and \K\
Kron photometry. 
Although many of the galaxy images have a considerable number of
stars superimposed on the images, comparison of the magnitudes derived from
this 
fairly automated algorithm agree well with the few known, independent
measurements.

\begin{figure} [ht]
\centerline {\epsfxsize=14cm \epsfbox[35 184 582 374]{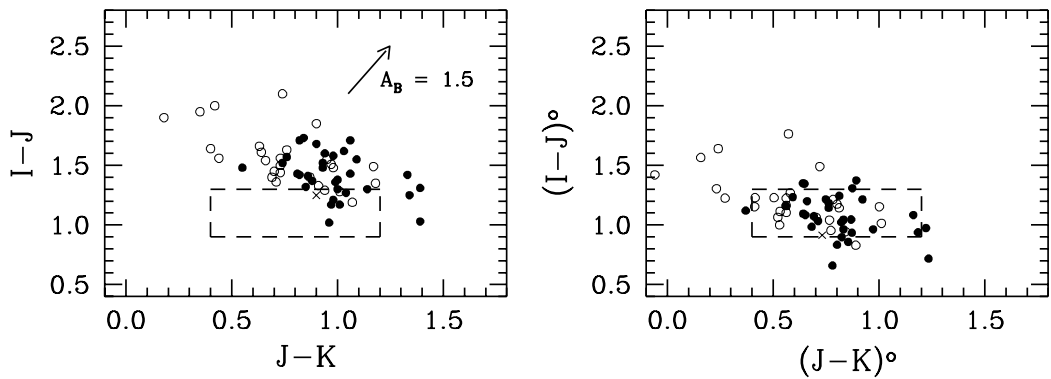}} 
\caption{The \ccd\ for the Norma cluster. Filled circles are early type
  galaxies, open circles spirals, crosses undetermined. The dashed box marks
  the 
  colour range of unobscured galaxies.  Observed colours are displayed in the
  left panel with the arrow indicating the expected mean offset due to
  extinction. The right panel shows extinction-corrected colours.}
\label{norijjkplot}
\end{figure}

The left panel of Fig.~\ref{norijjkplot} shows the colour\,--\,colour diagram
\ij\ versus \jk. As a comparison, the range in colours of galaxies in the
unobscured Hydra galaxy cluster as derived by Mamon (1997, in these
proceedings) is
indicated by the box. The displacement of the points agrees well with the path
of 
extinction (arrow) based on the mean extinction in the cluster of A$_B=1\fm5$
\cite{Sey}.

As the spread in extinction over the cluster area is large, we corrected the
colours of the galaxies individually for extinction using Mg$_2$-indices values
and interpolations according to the galactic \HI\ distribution. The
extinction-corrected values -- plotted in the right panel -- fill the dashed
box 
quite well, suggesting that our preliminary photometry is reasonably accurate
and that the shift in colour can be explained by the foreground extinction or,
vice-versa, that the NIR colours of obscured galaxies provide, in principle, an
independent way of mapping the extinction in the ZOA (see also \cite{gam2}).



\section{Conclusion }

Our pilot study illustrates the promises of using DENIS data for
extragalactic studies in the ZOA and mapping the Galactic extinction.
We expect to combine NIR photometry with
\HI\ data to derive distances independent of
redshift via the \tfr\ 
and therewith extend peculiar
velocity data to the ZOA.  
Moreover, the
discovery of `invisible' obscured galaxies in DENIS NIR images
opens the exciting possibility of comparing NIR galaxy lists from DENIS with
 galaxy lists obtained
from the ongoing systematic blind \HI\ galaxy survey
($v<13\,000$\kms ) at low latitudes ($|b|\le5\deg$) with the Multibeam
Receiver of the Parkes Radiotelescope \cite{MB}.



%
\vfill
\end{document}